\documentstyle[preprint,aps]{revtex}
\begin{document}
\draft

\title{Langevin Simulations of Two Dimensional Vortex
    Fluctuations:
    Anomalous Dynamics and a New $IV$-exponent }
\author{Kenneth Holmlund and Petter Minnhagen}
\address{Department of Theoretical Physics\\
    Ume{\aa} University\\
    901 87 Ume\aa, Sweden}
  \date{\today}
  \maketitle

\begin{abstract}
  The dynamics of two dimensional (2D) vortex fluctuations are
  investigated
  through simulations of the 2D Coulomb gas model in which
  vortices are
  represented by soft disks with logarithmic interactions.
  The simulations
  strongly support a recent suggestion that 2D  vortex fluctuations
  obey an
 intrinsic anomalous
 dynamics manifested in a long range $1/t$-tail
in the vortex correlations. A new  non-linear
 $IV$-exponent $a$, which is different from the commonly used AHNS
 exponent $a_{AHNS}$ and is given by $a=2a_{AHNS}-3$, is confirmed
 by the simulations. The results are discussed in the context of
 earlier simulations,
experiments and a phenomenological description.
\end{abstract}

\section{Introduction}
Superconducting films, 2D Josephson junctions, and $^4$He films
undergo
a Kosterlitz-Thouless(KT) type transition from the
superfluid to the
normal state.\cite{kosterlitz,minnhagen_rev} This transition is
driven by thermally created
vortex-antivortex pairs which start to unbind at the
transition.\cite{minnhagen_rev}
The high-$T_c$ materials can to some extent be regarded as
weakly coupled
superconducting planes, which raises the question to what
extent the 2D vortex fluctuations are important also for this new
class of materials.\cite{fischer} In order to assess
such questions it
is important to understand the properties of the vortex
fluctuations in the
pure 2D case. One motivation for the present investigation is to
gain such understanding.

The KT-transition is driven by the vortex fluctuations and
this means that
the large distance and long time behaviour in the
transition region
is dominated by the properties of the vortices.
The static properties
at the transition are described by the Kosterlitz
renormalization group
(RG) equations\cite{kosterlitz_rg} and are rather
well understood.\cite{halperin,minnhagen_rev} The properties
related to the dynamics of the vortices
constitute a much more open question.
We have in the present paper performed extensive simulations on a
simple dynamical model of vortex fluctuations in order to gain some
further insight.

So far the most widespread view on the dynamics of vortex
fluctuations derives from series of papers
by Ambegaokar et al.\cite{ambegaokar}
We refer to this as the AHNS phenomenology. Somewhat later
a variant was devised by one of us which we will
refer to as the MP-description.\cite{minnhagen_rev} These two
phenomenologies in fact give very different predictions.
Experimental
data seem to favor the
MP-description.\cite{minnhagen_rev,wallin,rogers}
Particularly clear experimental evidence for this was given by
Th\'{e}ron et al in case of a 2D array of Josephson
junctions.\cite{neuch}
The MP-description has also been clearly borne
out in computer simulations on 2D XY-type
models.\cite{westman,houlrik}
It has been argued that at the heart of
the MP-description is a long range $1/t$-tail in the vortex
correlations below the transition
temperature.\cite{minnhagen1} This
$1/t$-tail reflects an anomalous diffusion
of the vortex fluctuations
and is then the key to the difference
between the AHNS and the MP. In the present
paper we demonstrate that this anomalous diffusion is already a
property of the ``simplest'' possible dynamical model of vortex
fluctuations: A charge neutral system of positive and negative
particles with logarithmic particle interaction
and Langevin dynamics.

Another main prediction of the AHNS
is the exponent $a_{AHNS}$ of the
non-linear $IV$-characteristics
i.e. $V\propto I^{a_{AHNS}}$. However,
given the anomalous diffusion and the $1/t$-tail of the vortex
correlations, it has been argued that this prediction
is no longer correct.\cite{minnhagen1} A scaling
argument suggests that the
non-linear $IV$-exponent $a$ consistent with
the $1/t$-tail is given by $a=2a_{AHNS}-3$.\cite{minnhagen1} This
prediction is borne out to an excellent degree in the present
simulations.\cite{holmlund}

The content of the present paper is as follows: In section 2 we
describe the model and the simulation procedure. Sections 3-6
contain the results from the simulations. In section 3 we show
that the model and the simulations correctly reproduce the static
KT-transition {\em per se}. In particular we verify the power law
decay of the correlations for large distances
in the low temperature
phase. Good agreement with the predicted power law indices are
found. Then in section 4 we present the results for the
non-linear $IV$-exponent $a$ and verify the correctness of the new
scaling exponent. In section 5 we verify the $1/t$ tail
in the vortex correlations.
In particular we show how the decay of the
temporal correlations depend on the wavevector $k$ for
small $k$. Section 6 gives the
frequency dependence of the basic linear
response function describing
the coupling to an external electromagnetic field.
The results are
shown to be very well represented by
the functional form of the MP
description in the small $k$ limit.
We also show how the response function crosses over to a
a more Drude like behaviour as the wavevector $k$ is increased.
Finally
section 7 contains some concluding remarks.
\section{Model and Simulations}
In accordance with AHNS we assume that the dynamics of the vortices to
good approximation is described by the Langevin equation
\begin{equation}
\frac{d {\bf r}(t)}{dt}=\frac{D}{T}{\bf F}_{tot} (t) +{\bf \eta}(t)
\label{lang}
\end{equation}
where ${\bf r}$ is the position of a vortex
and ${\bf F}_{tot}$ is the total force acting on it due
to all the other particles as well as any externally imposed force,
$D$ is the diffusion constant, $T$ is the temperature
(unit system such that the Boltzmann constant $k_B=1$), and
${\bf \eta}$ is a random force obeying
\begin{equation}
  \langle \eta^\alpha (t) \eta^\beta (t')\rangle
  =2DT\delta_{\alpha\beta}\delta (t-t')
\label{noise}
\end{equation}
where $\alpha$ and $\beta$ denote the Cartesian components.
This equation describes the strong friction limit in which the
vortex motion is perpendicular to the applied external current and
should be a good approximation of a 2D
superconductor.\cite{ambegaokar,halperin,halperin1,minnhagen_rev}

According to the vortex-Coulomb gas analogy, the vortices can be
described as a gas of 2D Coulomb charges
with logarithmic interaction.\cite{minnhagen_rev}
The two possible vorticities $s\pm 1$ of a 2D vortex corresponds to
positive and negative Coulomb gas charge. The thermally created vortex
configurations have zero total vorticity which corresponds to a
neutral Coulomb gas.\cite{minnhagen_rev} In our model the Coulomb
gas charges are taken to be disks of extension $r_0$.
These disks correspond to the vortex cores and are such
that the force acting between two particles $i$ and $j$
with charges $s_i$ and $s_j$ respectively
(in units such that the charge is $s=\pm 1$) and
separated by the distance $r$ is given by
\begin{equation}
F_{ij}=s_is_j(\frac{1}{r}-\frac{1}{r_0}K_1(r/r_0))
\label{force}
\end{equation}
where $K_1$ is a modified Bessel function of order 1.
This means that the charge distribution of a
Coulomb gas particle is soft, which is in accordance
with the precise vortex-Coulomb gas particle analogy.\cite{nylen}
Consequently the force between two particles
vanishes for $r=0$ and is proportional to $1/r$ for $r>>r_0$.
Alternatively one may express the two particle interaction
corresponding to Eq.(\ref{force}) in terms of a potential $U(r)$
\begin{equation}
U(r)=-\ln (r/r_0) +K_0(r/r_0)
\label{pot}
\end{equation}
where $K_0$ is a modified Bessel function of order $0$,
\begin{equation}
  {\bf F}_{ij}=
  -s_is_j\frac{{\bf r}_{ij}}{r_{ij}}\frac{\partial}{\partial r_{ij}}
  U(r_{ij})
\end{equation}
and ${\bf r}_{ij}$ is the position vector from particle $i$ to
particle $j$. In the present paper length is in units of $r_0$ and
time in units of $t_0\equiv r_0^2/D$.

The simulations are performed for a fixed number of particles $N$ and
constant temperature $T$.
The particles are contained in a 2D quadratic box of
side length $L$ with periodic boundary conditions.
The numerical solutions were obtained by
discretizing time into small time steps
$\Delta t$ and introducing a random noise ${\bf \eta}(t)$
which acts independently
on each particle at each time step.
The Langevin equation (\ref{lang}) is then turned into a finite
difference equation for the particle system
\begin{equation}
  {\bf r}_i(t+\Delta t) = {\bf r}_i(t) +
  \Delta t \sum_{j=1,}^{N}{{\bf F}(r_{ij}(t))} +
\Delta t {\bf F}_{\mbox{ext}}(t) + {\bf \eta}_i(t),
\label{diff}
\end{equation}
where the indices $i$ and $j$ numerate the particles and the diffusion
constant $D$ has been absorbed into
the time scale and the random force.
${\bf F}(r_{ij}(t))=s_is_j\frac{{\bf r}_{ij}}{r_{ij}}F_{ij}$
is the force
acting at time $t$ on particle $i$ due to
particle $j$ and ${\bf F}_{ext}$ is any
external force.
The random force in Eq.(\ref{diff}) can
thus be treated as a random displacement vector
${\bf \eta}_i(t)$ which obeys (compare Eq.(\ref{noise}))
\begin{equation}
  \langle \eta^\alpha_i (t) \eta^\beta_j (t^\prime) \rangle =
  2 T \delta_{\alpha \beta}\delta_{ij} \delta (t-t^\prime),
\end{equation}
and is sampled from a Gaussian distribution.
This equation is then solved on the computer by
using a standard Euler integration method.\cite{num_ref}
For each temperature of interest the value
of $\Delta t$ was halved repeatedly until no dependence
of the time step could be monitored
(usually $\Delta t \approx 0.01\ t_0$).
The number of time steps needed for convergence is
usually $1-5\times 10^6$, but for the long
time correlation functions as many as
$15\times 10^6$ steps were needed in order
to obtain decently converged time tails.
In practice one has to strike a balance
between choosing $\Delta t$ small enough to
ensure that the equation of motion is correctly solved yet as
large as possible in order to achieve as large time
sequences as possible.
In practice we have been able to meet these
conditions below $T_c$ without too much problem.
However, just at and
slightly above $T_c$ this turned out to
be very computer time consuming.
Another practical problem is to keep track on the influence of the
boundary.
Here particular care has to be taken because the two particle
interaction is long range i.e. $U(r)\propto \ln r$ for large $r$.
To this
end we found it expedient to modify the interaction by a large
distance exponential cut off $\lambda_c$
which could then be varied in order to
check the dependence on the largest length scales. Thus
Eq.(\ref{force}) was modified into
\begin{equation}
  F_{ij}=s_is_j(\frac{1}{\lambda_c}K_1(r/\lambda_c)
  -\frac{1}{r_0}K_1(r/r_0))
\label{mod_force}
\end{equation}
corresponding to
\begin{equation}
U(r)=K_0(r/\lambda_c)-K_0(r_0/\lambda_c)-K_0(r/r_0)
\label{pot_mod}
\end{equation}
Typical parameters in the simulations are
$N=512$ and $L/r_0\approx 320$
which correspond to a
particle density $n=5\times 10^{-3}r_0^{-2 }$.
The ratio $\lambda_c/L=0.35$ turned out to be an efficient choice.
The size dependence of the results was checked by varying $L$ for
fixed $n$ and ratio $\lambda_c/L$. The size $L/r_0=320$ was in
practice large enough to avoid finite
size effects except very close to the
phase transition. In fact we found
that simulations on a $N=512$ system were
for practical purposes large enough for the
parameter range we are investigating.
However, to be on the safe side a fair amount
of the numerical data was obtained for
$N=1024$ and occasional checks
for $N=2048$ were also performed.

The Coulomb gas is often discussed in terms
of a fugacity variable $z$ where
$z^2/\Delta^2$ is the probability of creating
a dipole pair with particle separation $r_0$
and $\Delta$ is the phase space division for a particle.
This means
that in our model there exists a non trivial relation
between $nr_0^2$ and $z(nr_0^2,T)$.
However, in our present simulations $n$ and $T$
are the fundamental variables.

The basic correlation function
which we obtain from the simulations is
the Fourier transform of the charge density correlation function
$g({\bf r},t)$ defined as
\begin{equation}
\hat{g}({\bf k},t) = {1 \over L^2} \sum_{i,j}^{N}{s_i s_j e^{-i {\bf
      k}\cdot ({\bf r}_{i}(t)-{\bf r}_{j}(0))}}.
\end{equation}
In principle this function has a slight directional dependence on
${\bf k}$ due to our choice of periodic boundary on a quadratic box.
However, in practice our simulation
results are to good approximation
spherical symmetric so that $\hat{g}({\bf k},t)=\hat{g}(k,t)$.

The results are conveniently discussed in terms of
the complex frequency dependent dielectric constant ${1\over
\hat\epsilon(k,\omega)}$ of the Coulomb gas model
which is the basic response function.\cite{minnhagen_rev}
This is related to the correlation function $\hat{g}(k,t)$ by
\begin{equation}
  {1\over \hat\epsilon(k,\omega=0)}
  =Re\left[{1\over \hat\epsilon(k,\omega=0)}\right] =1
- {\hat{U} (k) \over T} \hat{g}(k,t=0)
\label{eps}
\end{equation}

\begin{equation}
  Re\left[{1\over \hat\epsilon(k,\omega)}\right]
  = Re\left[{1\over \hat\epsilon(k,\omega=0)}\right]
  + {\omega \hat{U}(k) \over T} \int_0^\infty{dt \sin{\omega t}
    \ \hat{g}(k,t)}
\label{Reps}
\end{equation}

\begin{equation}
  Im\left[{1\over \hat\epsilon(k,\omega)}\right]
  = - {\omega \hat{U}(k) \over T}
\int_0^\infty{dt \cos{\omega t} \ \hat{g}(k,t)}
\label{Ieps}
\end{equation}
The first equation (\ref{eps})
gives the static result which contains
the information on the thermodynamic KT-transition.
The following two,
(\ref{Reps}) and (\ref{Ieps}),
contain the information specific to the
dynamics of the model.

In addition to the linear response given by Eqs
(\ref{eps}-\ref{Ieps})
we calculate the non-linear response for the
case when the model is subject to an external force ${\bf
  F}_{ext}=s_i{\bf E}$
where ${\bf E}$ is constant in space and time. In this case we
calculate the average particle charge current $I_p$ per particle
\begin{equation}
  I_p={1\over N}\langle \sum^{N}_{i=1}s_i
  \frac{dr_i (t)}{dt}\rangle=
  {1\over N}\frac{D}{T}\langle \sum^{N}_{i=1}
  {\bf F}^{(i)}_{tot} (t) \rangle
\label{I_p}
\end{equation}
where the first equality is the definition of $I_p$
and the second follows directly from Eq.(\ref{lang}).
${\bf F}^{(i)}_{tot}$
is the total force acting on the particle $i$ and the
brackets $\langle \rangle$ denote a time average.

The results from these simulations are presented in the following
three sections.

\section{KT-transition.}
We will first focus on the static dielectric function
$1/ \hat{\epsilon}(k)\equiv 1/ \hat{\epsilon}(k,\omega=0)$ given by
Eq.(\ref{eps}). This function is related to the linearly screened two
particle interaction by
\begin{equation}
{1\over \hat\epsilon(k)}={\hat{U}_{eff}(k)\over \hat{U}(k)}
\label{eps_U}
\end{equation}
The ``bare'' interaction $\hat{U}(k)$
is in our case given by (compare Eq.(\ref{pot_mod}))
\begin{equation}
\hat{U}(k)={2\pi\over k^2+\lambda_c^{-2}}-{2\pi\over k^2+r_0^{-2}},
\end{equation}
provided $L=\infty$. In practice we use the numerical transform for
finite $L$ and periodic boundary conditions.
The linearly screened interaction is for small $k$
given by\cite{minnhagen_rev}
\begin{equation}
  \hat{U}_{eff}(k)=
  {1\over \tilde{\epsilon}}{2\pi\over k^2+\lambda^{-2} +O(k^4) }
\label{self}
\end{equation}
where $\lambda\leq \lambda_c$ is the screening length.
Consequently we expect that the static dielectric function for small
$k$ is of the form
\begin{equation}
{1\over \hat\epsilon(k)}={1\over \tilde{\epsilon}}{\hat{U}(k)\over
      k^2 +\lambda^{-2}}
\label{eps_fit}
\end{equation}
Figure 1 shows data for $1/\hat{\epsilon}(k)$ obtained from our
simulations for a sequence of temperatures at a fixed density $n$. The
filled circles
represent the data and the full curves are fits to
Eq.(\ref{eps_fit}). From these fits we
obtain $1/\tilde{\epsilon}$ and the screening
length due to free charges $\lambda_F$ defined as
$\lambda^{-2}_F=\lambda^{-2}-\lambda_c^{-2}$. These two quantities are
the key quantities describing the KT charge unbinding transition;
$1/\tilde{\epsilon}$ may be interpreted as describing the polarization
due to bound dipole pairs whereas $\lambda_F$ can be interpreted as
the Debye screening length related to the density $n_F$ of ``free''
charges
$\lambda_F^{-2}=2\pi n_F/ \tilde{\epsilon} T$.\cite{minnhagen_rev}
In the thermodynamic limit $L\propto \lambda_c\rightarrow\infty$ all
particles are bound into dipole pairs below the KT transition at
$T_c$ whereas above $T_c$ some pairs are broken.\cite{kosterlitz}
This means that $\lambda_F=\infty$ for $T<T_c$ and $\lambda_F<\infty$
for $T>T_c$.
In accordance with this figure 2
shows how $\lambda_F^{-2}$ obtained in our
simulations rapidly decreases as the KT transition is approached from
above. Precisely at the KT transition one has the condition
$\tilde{\epsilon}T_c=1/4$.\cite{kosterlitz,minnhagen_rev} This is
illustrated in figure 3 which shows $\tilde{\epsilon}$ as a function
of $T$ for a sequence of constant particle densities. One notes that
$\tilde{\epsilon}$ increases monotonously with increasing $T$ for low
temperatures, goes through a maximum and then decreases towards
$\tilde{\epsilon}=1$ for higher $T$. Roughly this means that first the
polarization due to bound pairs increases because the average
separation between the particles in
a bound pair increases and then the
polarization decreases because the number of bound pairs decreases due
to thermal pair breaking at higher temperatures. The full curve in
figure 3 corresponds to the condition
$\tilde{\epsilon}T=1/4$ and we use this as the determination of
$T_c$. This determination gives the phase transition line in the
$(n,T)$-plane, as shown in the insert of figure 3.
In the thermodynamic
limit $\tilde{\epsilon}$ has the critical
behaviour\cite{minnhagen_rev}
\begin{equation}
\tilde{\epsilon}(T)-\tilde{\epsilon}(T_c)\propto \pm\sqrt{|T-T_c|}
\label{e_crit}
\end{equation}
where + and - refer to above and below $T_c$. As seen in figure 3,
the weak singular behaviour implied by Eq.(\ref{e_crit}) cannot be
resolved by our present simulations. One notes, however, that the
determined $T_c$ is close to the inflection point of the numerically
obtained $\tilde{\epsilon}$-curve in accordance with
Eq.(\ref{e_crit}).
Associated with the weak singular behaviour of
Eq.(\ref{e_crit}) is a corresponding singular behaviour of
$\lambda_F$\cite{kosterlitz_rg}
\begin{equation}
\ln \lambda^{-2}_F\propto -{1\over \sqrt{T-Tc}}
\label{l_crit}
\end{equation}
as $T_c$ is approached from above.
In figure 4 $|\ln \lambda_F^{-2}|$ is plotted against $1/\sqrt{T-T_c}$
with $T_c$ determined from $\tilde{\epsilon}T_c=1/4$. As seen the
critical behaviour given by Eq.(\ref{l_crit})
is not discernible in the
simulations. However, this result is expected because
the true critical behaviour associated with Eq.(\ref{l_crit}) should
in practice be extremely hard to resolve as a
consequence of the extreme narrowness of the KT critical
region.\cite{region,olsson_crit}
In figure 4 we have also analyzed the data
with respect to Eq.(\ref{l_crit})
following a commonly used procedure in the context of superconducting
films and simulations on the 2D XY model:\cite{minnhagen_rev}
$|\ln \lambda_F^{-2}|$ is plotted against $1/\sqrt{T-T_c}$
where $T_c$ is a free parameter. As seen in figure 4 it is by this
procedure  possible to get a very good fit to Eq.(\ref{l_crit}). Such
fits are frequently claimed to be evidence for a KT critical
behaviour. However, as discussed in ref.\cite{olsson_crit} such fits
do usually not reflect a critical KT property per se,
but rather a property of the
2D Coulomb gas well outside the KT critical region. As is apparent
from figure 4, our present simulations are
consistent with this latter interpretation.

The low temperature phase displays a ``quasi'' 2D order in the sense
that the correlations for large distances fall off like power
laws.\cite{kosterlitz} In case of the charge density correlations we
have
\begin{equation}
g(r,t=0)\propto {1\over r^{x(T)}}
\label{g_pow}
\end{equation}
for $r>>r_0$ where\cite{kosterlitz_rg,minnhagen_rev}
\begin{equation}
x(T)={1\over \tilde{\epsilon}T}
\end{equation}
{}From a renormalization group (RG) point of view this means that each
$T\leq T_c$ corresponds to a fixed point in the
RG-flow.\cite{kosterlitz_rg}

The RG-flow is towards vanishing density $n$ so that  for
$T\leq T_c$ the line $(n=0,T)$ in the $(n,T)$-plane
is a line of fixed
points.\cite{kosterlitz} Each such fixed point corresponds to a
particular value of the critical index $x(T)$.
In figure 5 we show $g(r,0)$ as a function
of $r$ for a $T$ below $T_c$.
The function $g(r,0)$ was obtain by directly
measuring the charge correlations
as a function of distance for the
configurations generated
by the simulation.
The data for $g(r,0)$ is plotted as $\ln g(r,0)$
against $\ln r$ and according to
Eq.(\ref{g_pow}) the data should then
fall on a straight line with slope
$x(T)$ for large $r$. As seen in
figure 5 this prediction is borne out.
The broken straight lines in
figure 5 has the slopes given by
$1/\tilde{\epsilon}T$ where $\tilde{\epsilon}$
has been determined from
$1/\hat{\epsilon}(k,\omega =0)$ as described
in connection with figure 3. Thus the prediction
$x(T)=1/\tilde{\epsilon}T$ is supported to high degree by our
simulations.
The fact that the power law decay of the correlations
with distance and the power
law index come out correctly gives us
confidence in the present simulations.

\section{$IV$-exponent.}
Next we consider the non-linear response of the system when it is
subject to an external force ${\bf F}_{ext}=s_i{\bf E}$ where
${\bf E}$ is constant in space and time. This force generates a
particle charge current $I_p$. The charge current is in our
simulations obtained from Eq.(\ref{I_p}).
The prediction is that below
$T_c$ the generated charge
particle current is a power law in the
limit of small
magnitudes of the force\cite{ambegaokar}
\begin{equation}
I_p\propto F_{ext}^a
\label{power}
\end{equation}
In the context of a
2D superconductor the voltage $V$ is proportional
to the flux flow so that
$V\propto I_p$ whereas the force $F_{ext}$ is
proportional to the Lorentz force so
that $F_{ext}\propto I$ where $I$ is
the external current applied to the superconductor.
Thus in the
context of a 2D superconductor
Eq.(\ref{power}) corresponds to the
non-linear $IV$-characteristics
for small $I$ i.e. $V\propto I^a$.

The question we are addressing
with the present simulations is the value
of the exponent $a$.
There are two competing predictions: one is the
AHNS-prediction\cite{halperin,ambegaokar}
\begin{equation}
a_{AHNS}={1\over 2\tilde{\epsilon}T} +1
\label{ahns}
\end{equation}
and the other is a scaling prediction\cite{minnhagen1}
\begin{equation}
a={1\over \tilde{\epsilon}T }-1
\label{scaling}
\end{equation}

Figure 6 shows examples of
the $I_pF_{ext}$-characteristics obtained
from our simulations.
The data is plotted as $\ln I_p$ against $\ln
F_{ext}$ for a sequence of temperatures $T$. As is
apparent from the figure the data fall to
very good approximation on
straight lines for small $F_{ext}$, as predicted by
Eq.(\ref{power}). The slopes of these
lines give the values of the
exponent $a$. The full curves in figure 6 are
fits to the functional
form
\begin{equation}
I_p=CF_{ext}e^{-(a-1)K_0(BF_{ext})}
\label{special}
\end{equation}
where $a$ is the exponent and $B$ and $C$
are two constants. Fitting to
this functional form turned
out to be an expedient way of determining
the exponent $a$:
the $a$-values obtained by this fitting were the
same as the ones obtained
directly from the slope at small $F_{ext}$
but this latter procedure usually
required much more computer time.

A heuristic motivation for
Eq.(\ref{special}) goes as follows: the
particle current $I_p$ is proportional
to the density of free
particles $n_F$ and the force $F_{ext}$
i.e. $I_p\propto
F_{ext}n_F$. The free particle
density may be related to a self-energy
$U_{self}$ for the creation of
a free particle i.e. $\ln n_F \propto
U_{self}$.\cite{minnhagen_rev}
The self-energy corresponding to the
effective interaction in
Eq.(\ref{self}) is proportional to
$K_0(r_0/\lambda )$ where the
screening length $\lambda$ serves as an
effective cut off of the particle
interaction.\cite{minnhagen_rev} $F_{ext}^{-1}$ has
dimension of length and also serves
as an effective cut off of the
particle interaction.\cite{ambegaokar,minnhagen_rev}
Consequently one may expect
that whenever
$F_{ext}^{-1}<<\lambda$ the effective cut off in the
self-energy is proportional
to $F_{ext}^{-1}$. This argument suggests that
$\ln n_F \propto K_0(BF_{ext})$ and Eq.(\ref{special}) follows.

Figure 7 shows the obtained values for the exponent
$a$ as a function
of temperature $T$. These values are
in the figure compared to the two
competing predictions given
Eqs (\ref{ahns}) and (\ref{scaling}),
respectively.
In this comparison we use the values of
$\tilde{\epsilon}$ obtained
as described in section 2. As is
demonstrated by figure 7,
the scaling prediction given by
Eq.(\ref{scaling}) (full curve in the figure)
is borne out to high precision whereas the AHNS
prediction of Eq.(\ref{ahns})
(broken curve in the figure) clearly disagrees with the data.
One notes that the two predictions
agree precisely at the temperature
corresponding to $1/\tilde{\epsilon}T=4$
(crossing point between full
and broken curve in figure 7).
This corresponds to the critical
condition for the KT-transition and to the universal jump value
$a=3$ at $T_c$.\cite{nelson,halperin,ambegaokar}
Above $T_c$ there are free charges even in the limit
$F_{ext}=0$. Consequently one has
$I_p\propto n_F(F_{ext}=0)\neq 0$ for very
small $F_{ext}$ so that in
principle $a=1$ for $T>T_c$. Thus in
principle the exponents $a$ jumps
from 3 to 1 as $T_c$ is passed from
below. However, as seen in figure 7,
in practice the density of free
charges $n_F$ is dominated by the pair breaking
mechanism also above
$T_c$ for small $F_{ext}$.
This means that the exponent $a$
corresponding to pair breaking above
$T_c$ can in practice be
determined to very good precision,
as is apparent from from figure 7.
{}From figure 7 we infer that the pair
breaking exponent $a$ is to very
good approximation given by
the scaling prediction Eq.(\ref{scaling})
both below and above $T_c$.

The values of $a$ given in figure 7 are for
a fixed density $n$. In
general the exponent $a(T,n)$ is, of course,
a function of both $T$ and $n$.
Thus we can also test the prediction for $a$
as a function of $n$.
In figure 8 the data is plotted as a function
of $1/\tilde{\epsilon}T$
for four different densities.
The full straight line in figure 8
represents the scaling prediction of
Eq.(\ref{scaling}) and the broken straight
line the AHNS prediction of
Eq.(\ref{ahns}).
As seen in figure 8 the data falls clearly on the
full straight line representing the scaling prediction
for all the various densities. Thus we
conclude that the present simulations
strongly supports the scaling prediction.

An interesting observation in figure 8 is that the
exponent $a$ follows the scaling prediction
(given by the full
straight line) all the way down to
$a\approx 1$ close to $1/\tilde{\epsilon}T\approx
2$ at which point there is an
abrupt crossover to $a=1$. This suggests
an abrupt crossover behaviour at $T=1/2$
for small particle densities.
The 2D Coulomb gas model
has an equation of state which to
leading order in the particle density $n$ is given by\cite{hauge}
\begin{equation}
p=(T-\frac{1}{4})n \ \ \ \  {\rm for}\ \ \  T>\frac{1}{2}
\end{equation}
and
\begin{equation}
  p=\frac{1}{2}nT \ \ \ \ \ \ \ \ \  {\rm for}\ \ \  T<\frac{1}{2}
\end{equation}
where $p$ is the pressure.
For $T<1/2$ this equation of state can be
interpreted as the equation of state for an ideal gas of
non-interacting dipole pairs. This suggests
that the dominating part of
the gas
consists of dipole pairs in this small density limit.
For such a gas of dipole pairs free charges can be
generated by pair breaking
caused by an external force. On the other
hand for $T>1/2$ the equation
of state suggests a gas of free charges
with no bound dipole pairs.
This interpretation of the change of
behaviour of the equation of state at $T=1/2$ is in
accordance with the sharp crossover at $T=1/2$ of the exponent
$a$ which is seen in figure 8.

\section{Large $t$-dependence.}
In this section we focus on the large $t$-dependence of the charge
density correlations below $T_c$. Figure 9 shows our numerical data
for the Fourier transform $\hat{g}(k,t)$.
Our data suggest that the leading
small $k$ and large $t$-dependence is of the form
\begin{equation}
\hat{g}(k,t)\propto \frac{k^2e^{-const k^2 t}}{t}
\label{big-t}
\end{equation}
In order to establish this result we have in figure 9
plotted  the logarithm of  $t\hat{g}(k,t)/k^2$
against $t$ for a sequence of fixed values of $k$.
The form given by Eq.(\ref{big-t}) implies
that the data, when plotted in this way,
should for large  $t$ fall on straight lines. Furthermore
the slope of these lines should vanish as $k$
approaches zero. As seen in figure 9 these
features are very consistent with the data and
the data for the smallest $k$-values fall rather
nearly on horizontal lines.
The full straight lines in figure 9 are
least square fits to the data in
the region before too much noise sets
in. For the four largest
$k$-values in figure 9 such lines can be
determined without much uncertainty. The
corresponding slopes, together with estimated
uncertainties, are in figure 10 plotted against
$k^2$ (filled circles with error bars).
The broken straight line in figure 10 is a line
through the origin which is least square
fitted  to the determined slopes. The fact that
the slopes rather closely follow this line suggests
that the slopes are proportional
to $k^2$ for small $k$.
The broken straight lines in figure 9
have the slopes given by the open circles in
figure 10, i.e. they are the expected slopes for
these smaller $k$-values based on the
$k^2$-extrapolation of the slopes for the larger $k$-
values. As seen in figure 9 the broken lines
also fit rather well to the data, which lends
further support to the conclusion that
the slopes are indeed proportional to $k^2$ all the way
down to $k=0$. Thus we conclude that the data
in figures 9 and 10 support that the
small $k$ and large $t$ behaviour of
$\hat{g}(k,t)$ to good
approximation is given by Eq.(\ref{big-t}).
Simulations of the present type are of course
always hampered by limited system sizes and
time sequences. In particular we found that
the smaller the $k$-value the
harder it was to obtain a large $t$-value free of finite size
effects. E.g. the large $t$-part of
the two smallest $k$-values in
figure 9 remain somewhat uncertain.
Thus questions about logarithmic
corrections to the leading $t$-dependence
or non-leading terms appear
to be outside the limitation
of the present simulation precision.

In ref.\cite{minnhagen1} it was found that the function
$\lim_{k\rightarrow 0} \hat{g}(k,t)/k^2 \propto 1/t$
for large $t$ in
case of the 2D XY-model on a square
lattice with TDGL
(time-dependent Ginzburg-Landau type)-dynamics.
This result was in ref.\cite{minnhagen1}
associated with the vortex
fluctuations.
In the present paper we confirm this
conclusion by establishing the
result
directly in the Coulomb gas model with Langevin dynamics.
In addition we have obtained the leading
small $k$-dependence for large $t$.

The form given by Eq.(\ref{big-t})
implies that for large $t$ the dominant contribution
to $g(r,t)$ comes from the small $k$.
Thus we expect that the leading large
$t$ contribution to $g(r,t)$ is given by
\begin{equation}
  g(r,t)\propto \frac{1}{t}\int_0^{k_{{\rm max}}}
  dk^2k^2e^{-tk^2const}e^{i{\bf k}
    \cdot {\bf r}}\propto \frac{1}{t^3}
\end{equation}
for large $t$. The charge density is
obviously a conserved quantity in the
Coulomb gas model.
Thus a pile up of charge in one place
can only decay by diffusing
away.
Ordinary diffusion in 2D leads to $g(r,t)\propto 1/t$.
However, from our simulations of
the 2D Coulomb gas we conclude
that the long range interaction between
the particles changes this
result
into a more rapid decay
$g(r,t)\propto 1/t^3$ in the low temperature
phase $T<T_c$.
In the high temperature phase $T>T_c$
the screening length $\lambda$ is always
finite due to the presence of free charges.
Thus in this case the decay of the charge
density correlations are expected to decay
exponentially, where the decay is dominated
by a factor $\exp (-t\lambda^{-2} const)$. We
have not been able to explicitly verify this
result in the present simulations, since the
simulations are harder to converge in the high
temperature phase. We note that, since in the
small density limit the dipole pairs dominate
the response for $T<1/2$ (see the end of the
preceding section), one might likewise expect
that in practice the
behaviour $g(r,t)\propto 1/t^3$, which
we associate with the dipole pairs, also
dominates the response in a region somewhat above
$T_c$ ($T_c$ is always smaller than 1/4) for not
too large time scales. Thus one might
expect that the frequency response for small but
not too small frequencies are dominated by
the dipole pair response also in a region somewhat above $T_c$.

 The scaling prediction for the exponent $a$ given by
 Eq.(\ref{scaling}) was in ref.\cite{minnhagen1}
 based on the assumption that the charge
 density correlations $g(r,t)$
 can be associated with
 a scaling function $\lambda^{-z}\Phi
 (r\lambda^{-1}, t\lambda^{-z})$ where $z$
 is the dynamical exponent and $\lambda$
 is the screening length which
 diverges for any
 $T$ below $T_c$ in the limit $\lambda_c \rightarrow
 \infty$. Furthermore it was assumed
 that the scaling function $\Phi (x,y)$
 had the limits $\Phi (x,0)
 \propto x^{2-1/T\tilde{\epsilon}}$ for
 large  $x$ and $\Phi (0,y)\propto y^{-1}$ for large $y$.
 Consequently we can now infer that the relation
 between $g(r,t)$ and the invoked scaling form has to be
\begin{equation}
\lambda^{-z}\Phi (r\lambda^{-1}, t\lambda^{-z})\propto r^2t^2g(r,t)
\end{equation}
since $g(r_0,t)\propto t^{-3}$ for large $t$ and $g(r,t_0) \propto
r^{1/T\tilde{\epsilon}}$
(compare discussion in connection with figure 5, $t_0\equiv r_0^2/D$
is the microscopic time scale and $r_0$
is the microscopic size of a particle).

{}From Eqs (\ref{Reps}) and (\ref{Ieps})
one obtains the leading small
  $\omega$ dependence corresponding
  to Eq.(\ref{big-t}). Since according to Eq.(\ref{big-t})
  $F(t)\equiv \lim_{k\rightarrow 0}
  \hat{U}(k)\hat{g}(k,t)\propto
1/t$ for large $t$,
the leading small $\omega$-dependences are proportional to
\begin{equation}
\omega \int_0^\infty{dt \sin{\omega t} \  F(t)}\propto \omega
\int_0^\infty{dx \frac{\sin{x}}{x}}=\omega \ \frac{\pi}{2}
\label{small-omega}
\end{equation}
respectively,
\begin{equation}
-\omega \int_0^\infty{dt \cos{\omega t}\ F(t)}\propto -\omega
\int_{const}^{1/\omega} \frac{dt}{t}\propto \omega \ln{ \omega}
\end{equation}
and consequently
\begin{equation}
  Re\left[\frac{1}{ \hat\epsilon(0,\omega)}\right] -
  \frac{1}{\tilde{\epsilon}}=const \frac{\pi}{ 2 }\omega
\label{Re-small}
\end{equation}
\begin{equation}
  Im\left[\frac{1}{ \hat\epsilon(0,\omega)}\right]
  =const\ \omega\  \ln \omega
\label{Im-small}
\end{equation}
for small $\omega$. The constant $const$ depends on $T$ and has the
same value in the last two equations.

It is instructive to translate this result into the linear response
function corresponding
to the conductivity $\sigma (\omega )$ for a 2D superconductor.
The connection is $\sigma (\omega )\propto \left[-i\omega
\hat{\epsilon}(0,\omega )\right]^{-1}$.\cite{minnhagen_rev}
Consequently we have the leading small
$\omega$-dependence given by
\begin{equation}
Re \sigma (\omega )\propto -\ln |\omega |
\label{Rsigma}
\end{equation}
and
\begin{equation}
  Im \sigma (\omega )\propto \frac{constant}{\omega}
  + \frac{\pi}{2} {\rm sign}(\omega )
\label{Isigma}
\end{equation}
or equivalently
\begin{equation}
\sigma (\omega )\propto \frac{constant}{-i\omega}-\ln (-i\omega )
\label{rho-comp}
\end{equation}
For a 2D superconductor the dissipation for small $\omega$ are due to
thermally created vortices.
Hence Eq.(\ref{Rsigma}) suggests that for a 2D superconductor the real
part of the conductivity diverges logarithmically below $T_c$.

\section{Phenomenological description.}
As described above, we compute the Fourier transform of the charge
density correlation function
$\hat{g}(k,t)$ in our simulations. By aid
of Eqs (\ref{Reps}) and (\ref{Ieps})
we obtain the frequency dependence
of the dielectric function $1/\hat{\epsilon}(k,\omega )$. In figure 11
we present the result for
$Re\left[1/\hat{\epsilon}(k,\omega )\right]$
and $Im\left[1/\hat{\epsilon}(k,\omega)\right]$
for the smallest $k$ we
managed to converge ($k=0.039\ (r_0^{-1})$).
As is apparent from figure 9 we expect
that for such a small value of $k$ the real part of $1/\hat{\epsilon}$
should vanish linearly with $\omega$, as discussed in connection with
Eq.(\ref{Re-small}). For large $\omega$, on the other hand, the dipole
pairs have no time to respond so
that in this limit we expect that
$Re\left[1/\hat{\epsilon}(k,\omega )\right]
=1$.\cite{minnhagen_rev} Thus
we expect the real part for small enough $k$ to be of the form
$Re\left[1/\hat{\epsilon}(k,\omega )\right]\approx
Re\left[1/\hat{\epsilon}(0,\omega )\right]$
where
\begin{equation}
Re\left[\frac{1}{\hat{\epsilon}(0,\omega)}\right]
=\frac{1}{\hat{\epsilon} (0,0)}+
\left[1-\frac{1}{\hat{\epsilon}(0,0)}\right]
  \frac{\omega}{\omega + G(\omega )}
    \label{G}
    \end{equation}
    provided $0<G(\omega )<\infty$ and $G(\omega =0)=constant>0$. If
    $G(\omega )$ only depends weakly on $\omega $, we can approximate
    $G(\omega )$ by a positive constant $G(\omega)\approx
    \omega_0$. Using this approximation we can now obtain the
    corresponding approximation for the imaginary part by using the
    Kramers-Kronig relation. This leads to
    \begin{equation}
      Re\left[\frac{1}{\hat{\epsilon}(0,\omega )}\right]
      =\frac{1}{\hat{\epsilon}(0,0)}
      +\left[1-\frac{1}{\hat{\epsilon}(0,0)}\right]
      \frac{\omega}{\omega + \omega_0}
    \label{RMP}
    \end{equation}
and
    \begin{equation}
      Im\left[\frac{1}{\hat{\epsilon}(0,\omega)}\right]
      =-\left[1-\frac{1}{\hat{\epsilon}(0,0)}\right]
      \frac{2}{\pi}\frac{\omega\ \omega_0\
        \ln \omega}{\omega^2 -\omega_0^2}
    \label{IMP}
    \end{equation}
    One notes that Eqs (\ref{RMP}) and (\ref{IMP})
    correctly reduces to Eqs
    (\ref{Re-small}) and (\ref{Im-small})
    in the small $\omega $-limit.
The form of the $\omega$-dependence given by Eqs (\ref{RMP})
and (\ref{IMP}) is identical to the form given by the
MP-description.\cite{minnhagen_rev}
The MP-description was originally motivated
from a heuristical argument for the dipole
pair response.\cite{minnhagen_rev} As described
here, we can also view the MP-form as a simple interpolation
between two known limits. In figure 11 we have fitted Eqs
(\ref{RMP})and (\ref{IMP}) to
the data from the simulations by using the
two constants $\omega_0$ and
$\left[1-\frac{1}{\hat{\epsilon}(0,0)}\right]$
as fitting parameters. As seen from the figure, the
data is very well described by the MP-form.
{}From the fitting we
obtain $1/\hat{\epsilon}(0,0)=0.91$.
This is fairly close to the value
  obtained directly from the simulations
  $1/\hat{\epsilon}(k,0)
  =1/\tilde{\epsilon}\approx 0.92$ (compare
  figure 9, $k^{-2}\hat{g}(k,t
  =0)\propto 1-1/\hat{\epsilon}(k,\omega=0)$).
  The value obtained for $\omega_0$ from
  the fitting was $\omega_0\approx 0.36\pm 0.02$.
  This can be
  compared to $G(0)\approx 0.24$ which
  is obtained directly from
  the simulations: From the data in figure 9
  we get $\hat{g}(k,t)/k^2=A/t$
  for small $k$ and large $t$.
  By aid of Eq.(\ref{small-omega}) we then obtain
  the value of the $constant$ in front of
  the linear small $\omega $ dependence. Finally, by
using
$1/\hat{\epsilon}(k,0)$ from figure 1, we obtain $G(0)$ as
$G(0)\approx (1- 1/\hat{\epsilon}(k,0))/constant$. One notes
that $G(0)$ is somewhat smaller but of
the same magnitude as the
$\omega_0$ determined from the fitting
to the data, which supports the
assumption that $G(\omega )$ only
has a quite weak $\omega$-dependence.
Thus we conclude that Eqs (\ref{RMP} and (\ref{IMP}),
obtained from Eq.(\ref{G}) by using the
approximation $G(\omega )\approx \omega_0$,
gives a very good and quite
consistent description of the data from our present simulations.

So far we have focused on the $\omega$-dependence
in the small $k$-limit. Next we consider
how this $\omega$ dependence is changed
as $k$ is increased. Figure 12 shows the real
and imaginary part of $1/\hat{\epsilon}(k,\omega)$
as a function of $\omega$ for a sequence
of increasing $k$-values. As seen in figure 12
the imaginary part appears to be remarkably
independent of $k$ over the range of $k$-vales in the figure
($0.039\leq k\leq 0.31$),
whereas the real part decreases
significantly with increasing $k$.
However, since the real and imaginary part
are related by a Kramer-Kronig relation, the
change in the real part part must have a
corresponding change in the imaginary part. The
point is that the corresponding change in
the imaginary part is concentrated to small
frequencies.
Since the imaginary part is almost
independent of $k$ whereas the real part changes
significantly, one concludes
that Eqs (\ref{RMP}) and (\ref{IMP}) does
not describe the data as $k$ is increased, as
expected from the motivation of these equations.
In order to quantify this feature in a practical
way we focus on the change with $k$
precisely at the maximum of the imaginary part. As
seen in figure 12 this frequency is close
to $\omega\approx 0.36 $ and is approximately
independent of $k$ over the $k$ range
in the figure . Now the MP-form given by Eqs
(\ref{RMP}) and (\ref{IMP}) has the property
that, at the maximum frequency for the absolute
value of the imaginary part the, ratio between
the absolute value of the imaginary and real
part is precisely $2/\pi$. In figure 13 we have
plotted this ratio as a function of $k$. For small
$k$  this ratio approaches $2/\pi$ as expected
from Eqs (\ref{RMP}) and (\ref{IMP}).
However as $k$ increases this ratio also increases.
Ordinary diffusion corresponds to
\begin{equation}
  \frac{1}{\hat{\epsilon}(0,\omega)}
  =\frac{1}{\hat{\epsilon}(0,0)}+\left[1
  -\frac{1}{\hat{\epsilon}(0,0)}\right]
  \frac{\omega}{\omega + iDk^2}
    \label{drude}
    \end{equation}
    The real and imaginary part is for this
    case of the usual Drude form. The ratio between the
    absolute value of the imaginary and real part
    precisely at the maximum of the absolute value
    of the imaginary part is in the Drude case unity.
    It appears from figure 13 that the
    ratio approaches the ordinary Drude value unity as
    $k$ increases. This change of behaviour
    from MP to Drude is somewhat reminiscent of
    Eq.(\ref{big-t}) where the diffusion like factor
$\exp (-const\ k^2t)$ becomes more important with increasing $k$.

\section{Concluding remarks.}
The model investigated is a model of 2D vortex
fluctuations.
Consequently, the properties of this model should be
directly reflected in measurements on 2D
superconductors like superconducting films and 2D
Josephson junction arrays. In our simulations we verified that the
KT critical region is very narrow. For the resistance $R$ of a
superconductor the predicted KT-critical behaviour is\cite{halperin}
\begin{equation}
\ln R\propto \frac{1}{\sqrt{T^{CG}-T_c^{CG}}}
\label{Rs}
\end{equation}
  where $T^{CG}$ is the effective temperature variable for the
  vortices.\cite{minnhagen_rev}
  The narrowness of the critical region
  means that in practice the KT-critical
  behaviour probably cannot be resolved in resistance
measurements,
  as has been pointed out
  earlier.\cite{minnhagen_rev,region,olsson_crit} In our
  simulations we obtained $T_c$ to good precision.
  However, if we instead used
  $T_c$ as a free parameter,
  then we showed that the data could indeed be
  fitted to the KT-critical behaviour over
  several decades. This
  procedure of using $T_c$ as a free parameter
  and fitting to Eq. (\ref{Rs}) is the commonly
used way of establishing
  the KT-critical behaviour for the resistance $R$ of 2D
  superconductors.  It is frequently
  found that by treating $T_c$ as a
  free parameter the data can be
  nicely fitted to Eq.(\ref{Rs}), as was also
  the case for our simulations. However, in our
  simulations we also found that the $T_c$ from
  the fit was significantly different from (
  i.e.13\% lower than) the true $T_c$.
  Thus we conclude that these
  type of fits have in fact no direct
  bearing on the ``real'' critical
  behaviour.\cite{region,olsson_crit} It should
  also be noted that although the fitted $T_c$ for
  a superconductor can appear to be quite close to
  the true $T_c$ in real temperatures,  the
  corresponding difference in the
  effective vortex temperature variable
  $T^{CG}$ is usually much larger. The crucial point
  here is that our simulations suggest that, if
  the true $T_c^{CG}$ is used, then the data does
  not follow the functional form given by
 Eq. (\ref{Rs}).

 We also obtained the non-linear
 $IV$-exponent $a$ from the simulations
 and verified that
 $a=1/\tilde{\epsilon}T-1$\cite{comm}, as proposed in
 ref.\cite{minnhagen1} and which is
 quite different from the earlier
 AHNS-prediction\cite{halperin,ambegaokar}
 $a_{AHNS}$. However, precisely at $T_c$
 both prediction give $a=3$ since $a=2a_{AHNS}-3$.
 Below $T_c$ the new value is larger
 than the AHNS-value. In principle
 $a=1$ above $T_c$ due to usual flux
 flow resistance of free vortices.
 Nonetheless, we found that in practice
 a non-linear $IV$-exponent describing
 the pair breaking could also be
 determined to very good precision
 above $T_c$ up to roughly $2T_c$ at
 which point there was a rapid
 cross over to $a=1$. All the way up to
 roughly $2T_c$ we found
 $a=1/\tilde{\epsilon}T-1$ to very good approximation.

For a 2D superconductor the exponent $a$ is directly related to
$1/\tilde{\epsilon}
T^{CG}\propto \rho (T)/T$ where
the proportionality constant is a
combination of fundamental physical
constants and $\rho (T)$ is the 2D superfluid
density.\cite{minnhagen_rev}
Thus in order to test the prediction for $a$
one needs to know the temperature dependence
of $\rho (T)$. One way is to measure the complex impedance
$Z(T,\omega )$
since\cite{minnhagen_rev}
\begin{equation}
  Z(T,\omega )=
  -i\omega L_k (T) \hat{\epsilon} (k=0,\omega, T)\propto
\frac{T^{CG}}{T}\hat{\epsilon}(k=0,\omega, T)
\end{equation}
 where $L_k (T)$ is the sheet kinetic inductance and the
 proportionality factor is again just a
combination of fundamental physical constants.
Consequently, if $Z(T,\omega )$ is measured for
very small frequencies for a sample and the
non-linear IV-characteristics for the same sample
is measured to high precision, then the
prediction for $a$ for $T\leq T_c$ can be directly
put to experimental test. Alternatively, if
$Z(T,\omega )$ is measured for a small but
finite frequencies, then the $\omega =0$-limit
can be extracted by using
Eqs (\ref{RMP}) and (\ref{IMP}). For $T\geq T_c$
one needs both $L_k (T)$ and
$\tilde{\epsilon} (T)$. The sheet kinetic
inductance $L_k (T)$ can often be quite
well determined from the complex
impedance. The qualitative
behaviour of $\tilde{\epsilon} (T)$ is
clear from figure 3. It is also possible
to make a somewhat more quantitative
determination of $\tilde{\epsilon}$
above $T_c$ by starting from
$\hat{\epsilon}(k=0,\omega, T)$ at a
finite small frequency and fitting to the MP-
phenomenology.\cite{jonsson-un,houlrik}

We concluded from our model that 2D vortex fluctuations has a long
range $1/t$-tail in the vortex correlations below $T_c$. This means
that the conductivity $\sigma (\omega )$ is of the form given by
Eq.(\ref{rho-comp}). The crucial feature is that the next leading
term for small $\omega $ is a logarithm $\ln (-i\omega )$.
Such a logarithmic term is strongly supported by experiments on a 2D
array of Josephson junctions.\cite{neuch} These experiments are in
fact very well described by the MP-phenomenology.\cite{neuch} In
connection with these experiments there
has been other proposal for the
origin of this logarithm, e.g like vortex-spin wave
coupling\cite{beck} or as a specific
single vortex property of a proximity
coupled array\cite{koshunov}.
The point we are making here is that this logarithm is an {\em
  intrinsic} collective property of 2D vortex fluctuations which is
strongly linked to the long range logarithmic vortex interaction.

The general form of the frequency
dependence is given by Eq.(\ref{G}) which
reduces to the MP-form of Eqs (\ref{RMP}) and (\ref{IMP}) for
$G(\omega )=const $. The point
here is that, since there is no a priori
characteristic frequency scale other than the microscopic $t_0^{-1}$
one expects that $G(\omega )\approx G(0)$ as long as $\omega
<<t_0^{-1}$. We believe that the fact that the MP-form describes
a variety of experimental data as well as simulations very
well\cite{minnhagen_rev,wallin,rogers,neuch,westman,houlrik},
reflects this aspect:
the MP-form describes the dynamics for
frequencies much smaller than the basic
microscopic frequency scale and the functional
form is independent of the microscopic
details of  the dynamics. In this sense it
describes a universal behaviour of 2D vortex
fluctuations.

One may also note that $\omega_0(T)$ in Eqs (\ref{RMP}) and
(\ref{IMP}) depend on $T$. Thus the MP-form can be tested either for
fixed $T$ by varying $\omega $ or
for fixed $\omega $ by varying
$T$. This latter way is more common in experiments.
A particular
feature of the MP-form is the fact
that the ratio between the
imaginary and real part of the response
is $2/\pi$ precisely at the
dissipation peak.
The position of this peak is often quite well
defined in the experiments and peak
ratios close to $2/\pi$ have been
found both in experiments and
simulations.\cite{wallin,rogers,neuch,westman,houlrik}
In the present paper we
directly verify that the peak ratio $2\pi$ is an intrinsic
dynamical small frequency property of 2D vortex fluctuations.

\section*{Acknowledgment}
This work was supported by the Swedish Natural
Research Council through contract F-FU 04040-322. The authors are
indebted to Dr Dierk Bormann for stimulating discussions.

\newpage
\begin{figure}
\caption{\sl
\label{ekT.fig}
The static dielectric response function
$1/\hat{\epsilon}(k)$ plotted as a function of the wave
vector $k$ (in units of $2\pi/L$ where $L$ is the system size )
at a small constant particle density,
$n=0.005\ (r_0^{-2})$. The filled circles represent the data
from simulations
at $T$ $=$ $0.12$, $0.14$, $0.16$, $0.18$,
$0.20$, $0.23$, $0.26$, $0.29$, $0.32$, $0.35$,
$0.38$, $0.41$, $0.50$, from top to bottom, respectively.
The full curves are fits of equation \ref{eps_fit}.}
\end{figure}
\begin{figure}
\caption{\sl
  The inverse square of the screening length
  $\lambda_F^{-2}$ extracted from the data in
  figure 1 by using Eq.(\ref{eps_fit}). The open
  circles represent $\lambda_F^{-2}$ (in units of
  $(2\pi/L)^2$ and $n=0.005\ (r_0^{-2})$) plotted as a function
  of  temperature $T$.  The critical temperature $T_c$ is
  also shown ( $T_c$ is determined as described
  in connection with figure 3). The figure
  illustrates the charge unbinding transition:
  Below $T_c$ the charges are bound together in
  dipole pairs. These pairs start to unbind at
  $T_c$, resulting in a rapid increase in the density
  of free charges
  $n_F\propto \lambda_F^{-2}$ as $T$ passes $T_c$ from below.}
\end{figure}
\begin{figure}
\caption{\sl
\label{etiln.fig}
The dielectric constant $\tilde{\epsilon}$ extracted from the data in
figure 1 by using
Eq.(\ref{eps_fit}). The various symbols represent,
from bottom to top, the determined value of
$1/ \hat{\epsilon}(k)$ for the particle densities,
$n$ $=$ $0.001$, $0.005$, $0.01$, $0.025$,
$0.05$, $0.075$, $0.025$, $0.05$, $0.1$ ($r_0^{-2}$),
respectively. The broken curves are
guides to the eye. The full curve represent the KT
critical condition $\tilde{\epsilon} =
1/4T_c$. The full curve crosses the broken curves
close to the inflection points of the broken
curves as expected for the KT-transition. The crossing
points between the full curve and the
broken curves  gives the phase transition line in
the ($n$, $T$)-plane as shown in the insert.
The estimated error bars in the insert are due to numerical
uncertainties as well as finite size effects. It should be noted
that the uncertainties in the values of $T_c$ are very
small.}
\end{figure}
\begin{figure}
\caption{\sl
\label{test.fig}
Test of KT-criticality: The quantity $|\ln \lambda_F^{-2}|$
is plotted against $1/\protect\sqrt{T-T_c}$.
Plotted in this way the data should,
according to Eq.(\ref{l_crit}), fall on a straight line,
provided the data is in the critical region.
The data for $\lambda_F$ is the same as in figure
2. The filled circles represent the case
when the true $T_c$ is used ($T_c=0.215(7)$,
determined as described in connection with figure 3).
The broken curve is a guide to the eye.
As expected Eq.(\ref{l_crit}) does not describe
the data because of the unusual narrowness of
the KT critical region. The critical behaviour would
only show up closer to $T_c$ for much
larger values of $\lambda_F$ which cannot
be converged in the present simulations. If instead
$T_c$ is treated as a free parameter then data
can indeed be manipulated to fall on a
straight line. The open circles show the case for
$T_c=0.187$. In this case the data fall on a
straight line over several decades as
indicated by the dotted straight
line in the figure.}
\end{figure}
\begin{figure}
\caption{\sl
  The charge density correlation function
  $g(r, t=0)$ as a function of distance $r$ for
$T=0.18$, $n=0.005$ (circles), and $T=0.16$, $n=0.025$ (diamonds),
respectively. The data is plotted as
$|\log g(r,0)|$ against $\log r$ in
order to test the prediction $g(r,0)\propto r^{-1/\tilde{\epsilon}T}$
for large $r$. Plotted in this way the data should fall on straight
lines for large $r$ and this prediction is borne out. The broken lines
have the slopes given by $1/\tilde{\epsilon}T$ where
$\tilde{\epsilon}$ has been determined
as discussed in connection with
figure 3 and fit the data very well. Consequently the
prediction $g(r,0)\propto r^{-1/\tilde{\epsilon}T}$
is verified to high degree by the present simulations.}
\end{figure}
\begin{figure}
\caption{\sl
\label{iv.fig}
The charge current $I_p$ as a function of an
external force $F_{ext}=s_iE$
for the temperatures $T$ $=$ $0.12$, $0.14$,
$0.16$, $0.18$, $0.23$, $0.26$,
$0.29$, $0.50$ at constant density $n=0.005\ (r_0^{-2}$). The data
from the simulation are given by the filled
circles with error bars and is plotted as
$\log I_p$ against $\log F_{ext}$. The full curves are fits to
Eq.(\ref{special}). From these fits the exponents $a$ defined by
$I_p\propto F_{ext}^a$ are determined. Alternatively the exponent $a$
can be determined directly from the
slopes at small $F_{ext}$ (see text).}
\end{figure}
\begin{figure}
\caption{\sl
  The non-linear IV-exponent $a$ obtained from
  the data in figure 6. The exponent $a$ is plotted
as a function of temperature (filled circles with error bars).
The full curve is the scaling prediction Eq.(\ref{scaling}) and the
broken curve is the AHNS prediction Eq.(\ref{ahns}). The values of
$\tilde{\epsilon}$ needed to make the comparison
were determined as described in connection with
figure 3. The data strongly favors the scaling prediction.}
\end{figure}
\begin{figure}
\caption{\sl
  The IV-exponent obtained from
  the simulations at four different particle
  densities ($n=0.001$, $0.005$, $0.01$, and $0.025$ $(r_0^{-2}$)).
  The data
is plotted as $a$ against $1/\tilde{\epsilon}T$. Plotted in this way
the scaling prediction corresponds to the full straight line and the
AHNS prediction to the broken straight line. The vertical error bars
estimate the uncertainty in the value $a$ for a given $T$ and the
horizontal the uncertainty in $\tilde{\epsilon}$.
The data clearly verifies
the scaling prediction. Precisely at $T_c$ one has
$1/\tilde{\epsilon}T=4$ and both predictions give $a=3$. Note that $a$
in practice can be determined also above $T_c$ (see text) and clearly
follows the scaling prediction all the way upto
$1/\tilde{\epsilon}T\approx 2$ at which point there an abrupt cross
over to $a=1$.
}
\end{figure}
\begin{figure}
\caption{\sl
\label{gtt.fig}
The time dependence of the charge density correlation function $g$
below $T_c$ ($T=0.18$ and $n=0.005\ (r_0^{-2})$).
In order to verify the $t$-dependence
$\hat{g}(k,t)\propto k^2e^{-const k^2 t}/t$
given by Eq.(\ref{big-t}) the data is plotted
as $\log (t\hat{g}(k,t)/k^2)$ against $t$. The data
for large $t$ should then fall on straight lines where the slopes go
towards zero as $k$ is decreased. The data is given by the open
circles and the eight data sets correspond from top to bottom to the
$k$-values ${\bf k} = (1,1), (2,2), (3,3), (4,4), (5,5), (6,6), (7,7)$
and $(8,8)$ (in units of $2\pi/L$). The $t$-dependence of
Eq.(\ref{big-t}) is apparently borne out to fairly good approximation.
The full straight lines are fits to the data for
the larger $k$-values before too much noise sets
in. The broken straight lines have slopes extrapolated from the full
straight lines using Eq.(\ref{big-t}) as
described in connection with figure 10. Also the broken
straight lines fit the data fairly well.
}
\end{figure}
\begin{figure}
\caption{\sl
\label{slope.fig}
The $k$-dependence of the slopes in figure 9.
The slopes of the full straight
lines in figure 9 are plotted against $k^2$ (filled circles with error
bars). According to Eq.(\ref{big-t}) these slopes should extrapolate
linearly to zero. The broken straight line is a line through zero
which is fitted to the filled circles and shows that the slopes
apparently to good approximation are proportional to $k^2$.
The open circles give the expected slopes for some smaller $k$-values
and correspond to the broken straight lines in figure 9.
}
\end{figure}
\begin{figure}
\caption{\sl
\label{epswk11.fig}
The real (circles) and imaginary (diamonds) part of the frequency
dependent dielectric constant $1/\hat{\epsilon} (k,\omega )$ for a
small $k$ (the ${\bf k}$-vector is $(1,1)$ (in units of $2\pi/L$)
and the data is the same as the top
data set in figure 9). The full drawn curves are fits to the MP-form
given by Eqs (\ref{RMP}) and (\ref{IMP}) using the two constants
$\omega_0$ and $1-1/\hat{\epsilon} (0,0)$ as free parameters. The
MP-form describes the data very well.
}
\end{figure}
\begin{figure}
\caption{\sl
\label{allepsw.fig}
Real and imaginary parts of
$1/\hat{\epsilon} (k,\omega )$ for the same
parameters as in figure 9. The real part of
$1/\epsilon(k,\omega)$ is represented by the upper set of
curves. The amplitude decreases with increasing $k$ (the $k$-values
are the same as in figure 9).
The uppermost curve is the same data as was
shown to be well described by
the MP-form in figure 11. The $k$-dependence
of the imaginary part (lower set of curves) is by contrast very
small. Note that the absolute value of the imaginary part has a
maximum at $\omega \approx 0.36$ for all $k$-values in the figure.
The figure illustrates that the MP-form as
expected only describes the data in the
limit of small $k$.
}
\end{figure}
\begin{figure}
\caption{\sl
\label{peak.fig}
The peak ratio of the imaginary and real parts of
$1/\epsilon(k,\omega)$ as a function of $k$.
The peak ratios are obtained from the data shown in figure 12 and are
denoted by open circles. The broken curve is a guide to the eye.
For small $k$ the ratio
approaches the predicted MP value of $2/\pi$ whereas the Drude
ratio of unity is approached for large $k$.}
\end{figure}

\begin{thebibliography}{99}
\bibitem{kosterlitz} J.M. Kosterlitz and D.J Thouless,
  J. Phys C {\bf 5} L124 (1972);
J.M. Kosterlitz and D.J. Thouless, J.Phys. C {\bf 6}, 1181 (1973);V.L.
Berezinskii, Zh. Eksp.Teor.Fiz {\bf 61}, 1144 (1972) (Sov.Phys.JETP
{\bf 34}, 610, (1972)).
\bibitem{minnhagen_rev} For a review see e.g.
  P. Minnhagen, Rev. of Mod. Phys. {\bf 59}, 1001 (1987).
\bibitem{fischer} See e.g. K.H. Fischer,
  Physica C {\bf 210}, 179 (1993).
\bibitem{kosterlitz_rg} J.M. Kosterlitz,
  J. Phys. C {\bf 7}, 1046 (1974).
\bibitem{halperin} B.I. Halperin and D.R. Nelson,
  J. Low Temp. Phys. {\bf 36}, 599 (1979).
\bibitem{ambegaokar} V.~Ambegaokar, B.~I.~Halperin,
  D.~R.~Nelson, and E.~Siggia,
Phys.Rev.Lett. {\bf 40}, 783 (1978); V.~Ambegaokar,
B.~I.~Halperin, D.~R.~Nelson, and E.~Siggia, Phys.Rev. B {\bf 21},
1806 (1980);
V.~Ambegaokar and S.~Teitel, Phys. Rev B {\bf 19}, 1667 (1979).
\bibitem{wallin} M.~Wallin, Phys. Rev. B {\bf 41}, 6575 (1990).
\bibitem{rogers} C.~T.~Rogers, K.~E.~Myers,
  J.~N.~Eckstein, and I.~Bozovic,
                 Phys.Rev.Lett. {\bf 69}, 160 (1992).
\bibitem{neuch} R.~Th\'{e}ron, J.-B.~ Simond, Ch.~Leeman, H.~Beck,
P.~Martinoli, and P.~Minnhagen, Phys. Rev Lett. {\bf 71}, 1246 (1993).
\bibitem{westman} P. Minnhagen and O. Westman, Physica C {\bf 220},
347 (1994).
\bibitem{houlrik} J. Houlrik, A. Jonsson, and P. Minnhagen,
  Phys. Rev. B {\bf 50}, 3953 (1994).
\bibitem{minnhagen1} P. Minnhagen, O. Westman, A. Jonsson, and
  P. Olsson, Phys. Rev. Lett. {\bf 74}, 3672 (1995).
\bibitem{holmlund} A short account of these results have been
  published in K. Holmlund and P. Minnhagen,
  Proceedings of SUPNET '95, in press.
\bibitem{halperin1} B.I. Halperin in
 {\em Physics of Low-Dimensional Systems},
Proceedings of Kyoto Summer Institute, Sept. 1979,
edited by Y. Nagaoka and S. Hikami
 (Publication Office, Prog. Theor. Phys., Kyoto) p.53.
\bibitem{nylen} P. Minnhagen and M. Nyl\'{e}n, Phys. Rev. B {\bf 31},
  5768 (1985).
\bibitem{num_ref} A.Brass and H.J. Jensen, Phys. Rev. B {\bf 39},
  9587, (1989); D.L. Ermark, J. Chem. Phys. {\bf 62}, 4189, (1975).
\bibitem{region} P. Minnhagen and P. Olsson, Phys. Rev. B {\bf 45},
  10557 (1992).
\bibitem{olsson_crit} P. Olsson, Phys. Rev. B , in press (1995).
\bibitem{nelson} D.R. Nelson J.M. Kosterlitz, Phys. Rev. Lett {\bf
    39}, 1201, (1977); P. Minnhagen and G.G. Warren, Phys. Rev B {\bf
    24}, 2526, (1981).
\bibitem{hauge} E.H. Hauge and P.C. Hemmer, Physica Norvegica {\bf 5},
  209 (1971); P. Minnhagen, A. Rosengren and G. Grinstein, Phys. Rev B
  {\bf 18}, 1356 (1978).
\bibitem{comm} This has also been verified by H. Weber, M. Wallin and
H.J. Jensen, preprint (1995) and in ref.\cite{holmlund} above.
\bibitem{jonsson-un} A. Jonsson and P. Minnhagen, unpublished.
\bibitem{beck} H. Beck, Phys. Rev. B {\bf 49}, 6153 (1994).
\bibitem{koshunov} S.E. Koshunov, Phys. Rev. B {\bf 50}, 13616 (1994).
\end{thebibliography}
\end{document}